\documentclass[preprint2]{aastex61}

\usepackage{amsmath}
\usepackage{hyperref}
\usepackage{listings}
\usepackage{color}
\usepackage{float}
\usepackage{xspace}
\usepackage{inconsolata}

\usepackage[footnote,printonlyused]{acronym}
\usepackage{scrextend}
\deffootnote[1.0em]{1em}{0em}{\textsuperscript{\thefootnotemark}\,\enskip}

\newcommand{\astroplan}{\texttt{astroplan}\xspace}
\newcommand{\astropy}{\texttt{astropy}\xspace}
\newcommand{\pyephem}{\texttt{pyephem}\xspace}

\definecolor{dkgreen}{rgb}{0,0.6,0}
\definecolor{gray}{rgb}{0.5,0.5,0.5}
\definecolor{mauve}{rgb}{0.58,0,0.82}

\begin{document}

\title{astroplan: An Open Source Observation Planning Package in Python}
\shorttitle{astroplan: Observation Planning Package in Python}
\shortauthors{Morris et al.}

\author[0000-0003-2528-3409]{Brett M. Morris}
\affiliation{Astronomy Department, University of Washington, Seattle, WA 98195, USA}

\author{Erik Tollerud}
\affiliation{Space Telescope Science Institute,  3700 San Martin Dr, Baltimore, MD 21211, USA}

\author{Brigitta Sip\H{o}cz}
\affiliation{Cambridge, UK}

\author{Christoph Deil}
\affiliation{Max Planck Institute for Nuclear Physics, Saupfercheckweg 1, 69117 Heidelberg, Germany}

\author{Stephanie T. Douglas}
\affiliation{Harvard-Smithsonian Center for Astrophysics, 60 Garden St, Cambridge, MA 02138}

\author{Jazmin Berlanga Medina} 
\affiliation{Imagination Station, Lafayette, IN, USA}

\author{Karl Vyhmeister}
\affiliation{California Institute of Technology, Pasadena, CA 91125, USA}

\author{Toby R. Smith}
\affiliation{Astronomy Department, University of Washington, Seattle, WA 98195, USA}

\author{Stuart Littlefair}
\affiliation{University of Sheffield, Sheffield S10 2TN, UK}

\author{Adrian M. Price-Whelan}
\affiliation{Department of Astrophysical Sciences, Princeton University, Princeton, NJ 08544, USA}

\author{Wilfred T. Gee}
\affiliation{Department of Physics and Astronomy, Macquarie University, NSW 2109, Australia}

\author{Eric Jeschke}
\affiliation{Subaru Telescope, National Astronomical Observatory of Japan, 650 North A'ohoku Place, Hilo, HI 96720, USA}

\email{bmmorris@uw.edu}

\begin{abstract}
We present \astroplan\ --- an open source, open development, Astropy affiliated package for ground-based observation planning and scheduling in Python. \astroplan is designed to provide efficient access to common observational quantities such as celestial rise, set, and meridian transit times and simple transformations from sky coordinates to altitude-azimuth coordinates without requiring a detailed understanding of \astropy's implementation of coordinate systems. \astroplan provides convenience functions to generate common observational plots such as airmass and parallactic angle as a function of time, along with basic sky (finder) charts. Users can determine whether or not a target is observable given a variety of observing constraints, such as airmass limits, time ranges, Moon illumination/separation ranges, and more. A selection of observation schedulers are included which divide observing time among a list of targets, given observing constraints on those targets. Contributions to the source code from the community are welcome.
\end{abstract}


\section{Introduction}

The Astropy Project is a community effort to develop a common core package for astronomy in Python, and to foster an ecosystem of interoperable astronomy packages. The \astropy core package contains all of the machinery necessary for computing whether or not a given object is observable from a location on the Earth at specified times. It defines an object-oriented framework for specifying times, and coordinates on the sky and Earth. In this paper, we assume that the reader has some familiarity with the tools available in \astropy, see \citet{Astropy2013} or the online documentation\footnote{\url{http://docs.astropy.org}}. 

There are several practical algorithms useful for observation planning that are not included in \astropy. Some questions that users may seek to answer using \astropy would require substantial effort, such as: ``is this star currently above $30^\circ$ altitude from the Apache Point Observatory?'' or ``what time is astronomical twilight this evening on Mauna Kea?''

\astroplan is an Astropy affiliated package for ground-based observation planning and scheduling, which provides functionality for answering these questions. It is a pure-Python package that provides an efficient application programming interface (API) for quick access to common observational calculations, while using the full accuracy and precision of \astropy under-the-hood to handle the sky and time coordinate transformations. 

The most similar existing Python software that can be used to plan observations is \pyephem \citep{pyephem}. \astroplan is different from \pyephem in a few fundamental ways. \astroplan provides support for computing the positions of the Sun, Moon, stars, and major planets. It uses \astropy's modern and more accurate IAU2000/2006 methods and NASA's DE430 planetary ephemeris. \astroplan is built around the \astropy objects which specify times and coordinates. \astroplan users can use the extensively documented and constantly improving \astropy framework for specifying times and coordinates. \pyephem uses package-specific implementations of times and coordinates that are not cross-compatible with packages in the Astropy Project ecosystem. \pyephem supports the Sun, Moon, stars, major planets, asteroids and comets, and uses the older IAU1976/1980 precession/nutation methods, and VSOP87 planetary ephemerides.

Here we briefly outline some key features of \astroplan version 0.4, and the design decisions that went into making them. In Section~\ref{sec:api} we outline the \astroplan API, and in Section~\ref{sec:docs} we direct the reader to the online documentation, and resources for teaching and learning \astroplan.

\section{API} \label{sec:api} 

\subsection{Basic operations}

We begin by defining the \texttt{Observer} object, which specifies the location of an observer on the Earth. Most of the major observatories included in IRAF \citep{IRAF} are accessible by name in \astroplan via the \texttt{at\_site} class method:
\begin{lstlisting}[caption=Define a common observer][H]
>>> from astroplan import Observer
>>> keck = Observer.at_site("Keck") 
>>> apo = Observer.at_site("Apache Point")
>>> print(apo)
<Observer: name='Apache Point',
    location (lon, lat, el)=(-105.822 deg, 32.78 deg, 2798 m),
    timezone=<UTC>>
\end{lstlisting}
An observer can be located anywhere on the Earth with use of \astropy's \texttt{EarthLocation} object.
\begin{lstlisting}[caption=Define a custom observer][H]
>>> from astropy.coordinates import EarthLocation
>>> import astropy.units as u

>>> longitude = '-155d28m48.900s'
>>> latitude = '+19d49m42.600s'
>>> elevation = 4163 * u.meter
>>> location = EarthLocation.from_geodetic(longitude, latitude, elevation)
>>> observer = Observer(name='Subaru', location=location)
\end{lstlisting}
In order to account for atmospheric refraction in different environments, several atmospheric parameters can be described on the \texttt{Observer} object, including the atmospheric pressure, temperature and relative humidity. 

Targets with fixed celestial coordinates are described by \texttt{FixedTarget} objects, which contain their coordinate and name:
\begin{lstlisting}[caption=Define a fixed celestial target][H]
>>> from astroplan import FixedTarget
>>> from astropy.coordinates import SkyCoord

>>> sirius = FixedTarget.from_name("Sirius")
>>> vega_coord = SkyCoord(ra="18h36m56s", dec="+38d47m01s")
>>> vega = FixedTarget(coord=vega_coord, name="Vega")
\end{lstlisting}
The \texttt{from\_name} class method uses tools from \texttt{astropy.coordinates} to query Simbad, NED, and VizieR for target coordinates by name through the Sesame Name Resolver \citep{Schaaff2004}. Non-fixed targets apart from the Sun and Moon are not implemented in astroplan at the time of writing, and community contributions for supporting minor bodies are welcome.

Rise and set times are the cornerstone computations of observation planning. \astroplan computes the rise and set times of an object by transforming the sky coordinates of the object (e.g.~ICRS, galactic, etc.) into a grid of altitude-azimuth coordinates for that target as seen by an observer at a specific location on the Earth, at 10 minute intervals over a 24 hour period. The rise or set time is then computed by linear interpolation between the two coordinates nearest to zero. The meridian/anti-meridian transit time is computed similarly; it takes a numerical derivative of the altitudes before searching for the appropriate zero crossing. The user can also define a rise or set horizon other than $0^\circ$ altitude, which is useful for observatories with non-zero altitude limits. 

We chose to compute rise and set times with a grid-search to maximize accuracy, rather than speed. In particular, we sought to preserve the \astropy altitude-azimuth coordinate transformation which accounts for atmospheric refraction. 

Convenience methods are included to compute the altitude-azimuth coordinates of a target at a given time, and the times of rise, set, meridian and anti-meridian transit:
\begin{lstlisting}[caption=Find target altitude/azimuth and rise time][H]
>>> from astropy.time import Time
>>> observing_time = Time("2017-01-01 05:23:45", scale="utc")
>>> apo.altaz(observing_time, sirius)
<SkyCoord ...: (az, alt) in deg
    ( 150.06353683,  34.88938344)>
>>> apo.target_rise_time(observing_time, sirius)
<Time object: scale='utc' format='jd' value=2457754.5765391565>
\end{lstlisting}
Times can be defined in a variety of scales using \astropy \texttt{Time} objects, including UTC, TAI, TCB, TCG, TDB, TT, UT1.

The sky coordinates of the major Solar System bodies are computed using the \texttt{jplephem} package, which provides an API for querying JPL's Satellite Planet Kernel files. The methods for querying the positions of Solar System bodies were originally developed for \astroplan, and have since been moved into the \texttt{astropy.coordinates} package. 

\begin{figure}
\centering
\includegraphics[scale=0.5]{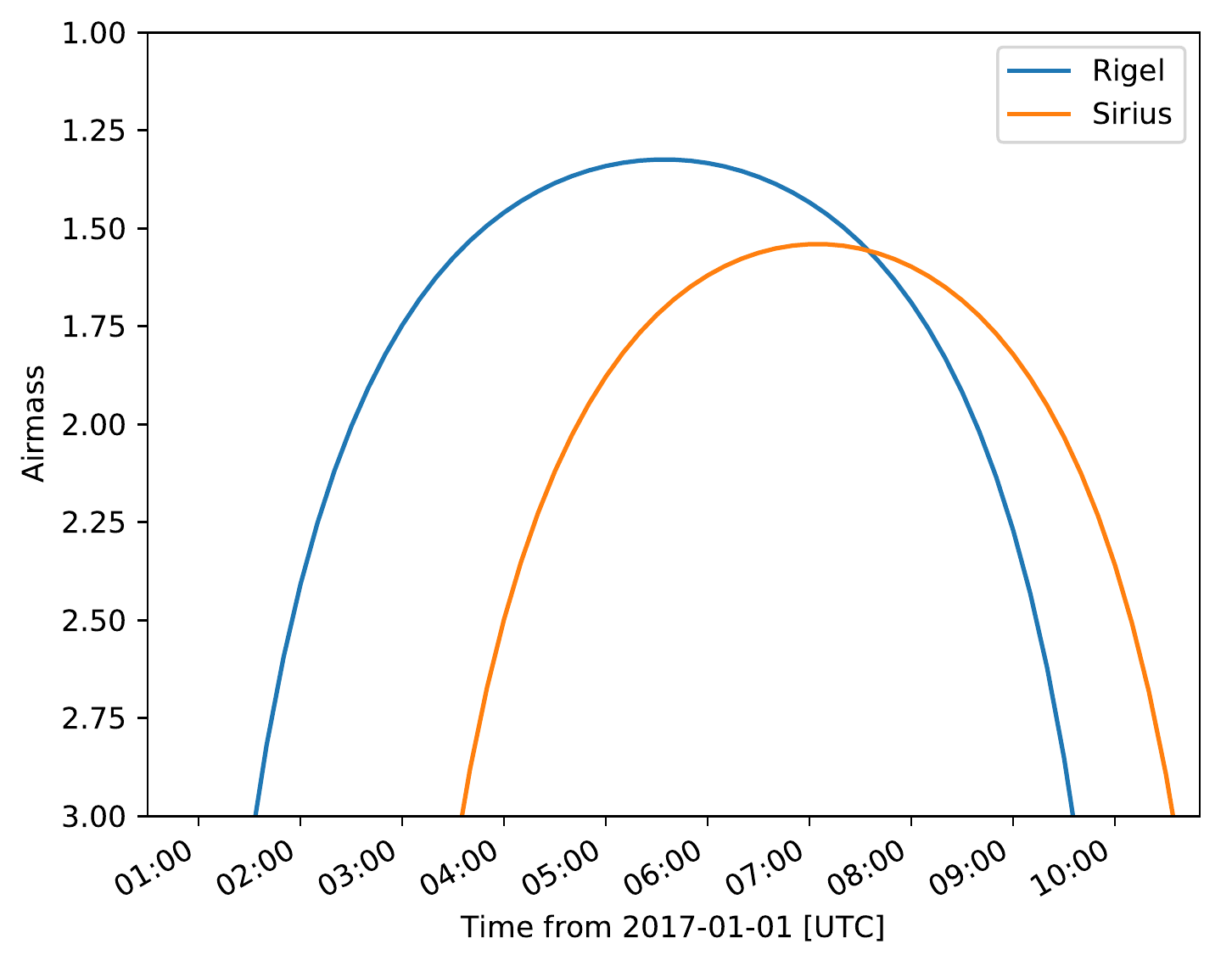}
\includegraphics[scale=0.5]{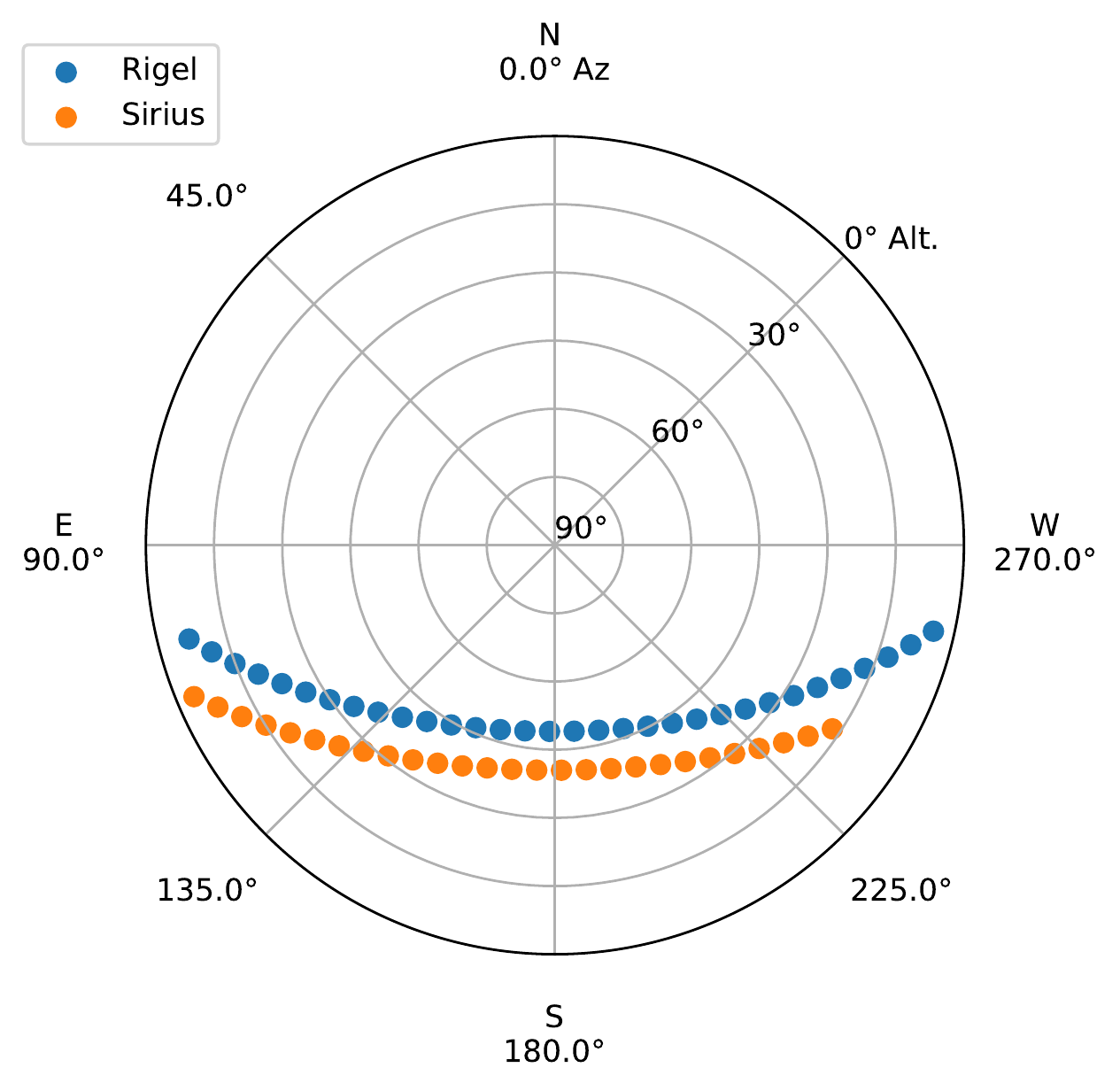}
\caption{An airmass plot and a sky chart for observing Sirius and Rigel from Apache Point Observatory, made with the \texttt{astroplan.plots} methods \texttt{plot\_airmass} and \texttt{plot\_sky}. The underlying altitude/azimuth calculation powered by \texttt{astropy.coordinates} includes atmospheric refraction.}
\label{fig:plots}
\end{figure}

Common plots are accessible through the \texttt{astroplan.plots} module -- see Figure~\ref{fig:plots} for a few examples. There are many more example plots, and the source code that generates them, available in the online documentation\footnote{\url{https://astroplan.readthedocs.io/en/stable/tutorials/plots.html}}.

\subsection{Observing Constraints} \label{sec:constraints}

Planning astronomical observations often requires an observer to determine whether or not a celestial object is observable given a list of observing constraints. \astroplan contains a generic framework for defining observing constraints, and computing the ``observability'' of a list of targets given those constraints.

For example, suppose an observer is planning to observe low-mass stars in Praesepe in the optical and infrared from the W.M.~Keck Observatory. The constraints imposed by the telescope and science case require all observations to occur: (i) between astronomical twilights; (ii) while the Moon is separated from Praesepe by at least $45^\circ$; and (iii) while Praesepe is above the lower elevation limit of Keck I, about $33^\circ$. These observing constraints can be specified with the \texttt{AtNightConstraint}, \texttt{MoonSeparationConstraint}, and \texttt{AltitudeConstraint} objects. We demonstrate this use case with \astroplan in a long code example in Section~\ref{sec:constraints_example}.

Other built-in constraints allow users to specify acceptable ranges of: Moon illuminations, airmass, Sun separations (e.g.,~for non-optical observations), and local times. The observing constraint classes take as input: targets, times and an observer; and the constraints return boolean matrices indicating whether or not those targets are observable at each time. 

The constraints framework is modular and written to be extensible. Users can implement their own constraints for a particular observatory or science case by following a tutorial in the online documentation\footnote{\url{https://astroplan.readthedocs.io/en/stable/tutorials/constraints.html\#user-defined-constraints}} to produce constraint objects which are compatible with the \astroplan scheduling framework.

\subsection{Transiting exoplanets and eclipsing binaries}

The \texttt{astroplan.periodic} module contains a framework for defining systems with periodic events, such as exoplanets and binaries. There are specialized classes for eclipsing systems, such as eclipsing binaries (EBs) and transiting exoplanets. The module makes use of the generic terms ``primary eclipse'' and ``secondary eclipse'', where the primary eclipse is a ``transit'' in the case of exoplanets. There are convenience functions for computing the next primary or secondary eclipses of an exoplanet or EB, or as well as computing ingress and egress times of the next primary or secondary eclipse.
\begin{lstlisting}[caption=Find upcoming exoplanet transit times][H]
>>> from astropy.time import Time
>>> import astropy.units as u
>>> from astroplan import EclipsingSystem

>>> epoch = Time(2452826.628514, format='jd', scale='utc')
>>> period = 3.52474859 * u.day
>>> duration = 0.1277 * u.day
>>> now = Time('2016-09-15')

>>> hd209458 = EclipsingSystem(epoch, period, duration)
>>> print(hd209458.next_primary_eclipse_time(now, n_eclipses=3))
['2016-09-16 23:37:44.154' 
 '2016-09-20 12:13:22.433' 
 '2016-09-24 00:49:00.711']
\end{lstlisting}

There are also complementary methods in the \texttt{constraints} module for use with the \texttt{periodic} system framework. Users can determine which eclipse events are observable from an observatory with a list of constraints. We include a brief tutorial for using the \texttt{periodic} module with queries from online exoplanet parameter databases in Appendix~\ref{sec:eb}.

\subsection{Scheduling Observations}

The scheduling framework enables users to define \textit{observing blocks}, which denote an observation of a target or group of targets for an amount of time in a particular instrument configuration. Each observing block can be assigned a numerical \textit{priority}, which by convention spans the range [0, 1] where zero is low priority. Priorities can be assigned by an observer based on which potential observations are most important to them to get scheduled. A set of observing blocks gets assigned a \textit{rank}, which for example, might be the rank a proposal receives from a telescope time allocation committee (TAC). 

Each observing block has a list of associated \textit{constraints}. We compute a \textit{score} for each constraint on an observing block, which can be a boolean or float in the range [0, 1] where zero is unfavorable. For example, the score computed from an airmass constraint will be highest when the airmass is low, while the score computed from an altitude constraint will be highest when the altitude is high. Other constraints, like the \texttt{AtNightConstraint}, yield boolean scores.

These scored observing blocks can be assigned to time slots by a \textit{scheduler}, which chooses the order for which observing blocks get scheduled first, and the times to assign them. Each scheduler creates an observing schedule based on one of several strategies for filling time slots with observing blocks. As of \astroplan version 0.4, there are two schedulers implemented: the sequential and priority schedulers. 

The \textit{sequential scheduler} begins by selecting the best-scored observing block at the beginning of the observing time. It then continues to choose the next best-scored block for the next observation, until all available observing time is allocated, or all observing blocks have been allocated. 

The \textit{priority scheduler} takes a prioritized list of observing blocks. The priority for each observing block could be assigned by an observatory TAC for example, or by an individual observer who needs to schedule their observations given their scientific priorities. The scheduler will first allocate the highest priority observing block to the best-scored time slot for that observing block, and then schedule the next priority block at its best time, etc. 

The two schedulers presently implemented are most useful for planning an individual observer's observations; a complete example is available in Appendix~\ref{sec:scheduleapp}. We intend to continue to develop the scheduling module to support queue scheduling for observatories with many observing programs. A wide range of strategies exist for planning observations, however, so the code for the schedulers is adaptable for users to adopt to other strategies either via subclassing or creating new scheduler classes.  The package welcomes contributions of this sort from the community.

\subsection{Testing \& Development}

\astroplan has an extensive testing suite. In addition to simple unit tests which check that sensible inputs yield sensible outputs, there are also many tests which compare the accuracy of \astroplan outputs. The tests are executed remotely whenever changes are made to the source code or documentation within the \astroplan repository. The \astroplan outputs are commonly compared against outputs from the independent python ephemeris package \pyephem \citep{pyephem}. The difference in rise and set times with \astroplan and \pyephem is always $<8$ minutes (with atmospheric refraction), and the differences are probably attributable to intrinsically different interpretations of these times.

Contributions to the package from the community are welcome. The source code is hosted on GitHub\footnote{GitHub: \url{https://github.com/astropy/astroplan}, \linebreak static Zenodo archive: \url{https://doi.org/10.5281/zenodo.1035883}}, where users can contribute new features. \astroplan follows the open development model refined by \astropy, and many tutorials on contributing to the source code of either package are available in the \astropy documentation\footnote{\url{http://docs.astropy.org/en/stable/development/workflow/development_workflow.html}}.

\section{Documentation} \label{sec:docs}

\subsection{Online Documentation} 

Detailed, tested, living documentation for \astroplan is available online via Read the Docs\footnote{\url{http://astroplan.readthedocs.io/}}. This paper is intended as a brief introduction to \astroplan's core functionality and the algorithms used throughout the package, so we refer the reader to the online documentation for the complete API description, and complete tutorials for each module with examples.

\subsection{\astroplan in the classroom} \label{sec:teaching}

\astroplan is incorporated into the curriculum for undergraduate majors in astronomy at the University of Washington, in the ``Introduction to Programming for Astronomical Applications'' course. The lesson plan on observing with Python is built around the task of planning astronomical observations. Along the way, it guides students through using the time, coordinate and quantity objects of \astropy, building up to their combined use in observation planning with \astroplan. Jupyter notebooks guiding students through these lessons are freely available online\footnote{\url{https://github.com/UWashington-Astro300/astroplan-in-the-classroom}}.

\section{Summary}

\astroplan is a pure-Python, open source, Astropy affiliated package for observation planning and scheduling. It provides methods for computing common observational quantities such as target rise, set, transit times; and it specifies a framework for testing the ``observability'' of targets given observing constraints. 

\software{\texttt{astroplan} \citep{astroplan}, \texttt{ipython} \citep{ipython}, \texttt{numpy} \citep{VanDerWalt2011}, \texttt{scipy} \citep{scipy},  \texttt{matplotlib} \citep{matplotlib}, \texttt{astropy} \citep{Astropy2013}, \pyephem \citep{pyephem}, \texttt{jplephem}\footnote{\url{https://github.com/brandon-rhodes/python-jplephem/releases/tag/v2.6}}}

\acknowledgements

B.M.M., J.B.M. and K.V. gratefully acknowledge support from the Google Summer of Code program in 2015 and 2016. B.M.M. acknowledges financial support from the Python Software Foundation; and from the University of Washington eScience Institute, with funding from the Gordon and Betty Moore Foundation and the Alfred P. Sloan Foundation. We thank Eric Agol and Suzanne Hawley for supporting B.M.M.~to devote some PhD thesis time towards developing and maintaining \astroplan.

This research has made use of NASA's Astrophysics Data System. This research has made use of the SIMBAD database, operated at CDS, Strasbourg, France \citep{Wenger2000}.

\appendix

We outline here some in-depth code examples which demonstrate a few intended use cases for \astroplan. We again encourage the reader to visit the online documentation described in Section~\ref{sec:docs} for many example inputs and outputs.

\section{Observing constraints} \label{sec:constraints_example}
In Section~\ref{sec:constraints}, we outlined a list of example observing constraints, which we might like to evaluate at various times with \astroplan. We will observe Praesepe from Keck Observatory, and we are setting the following constraints: (i) observe between astronomical twilights; (ii) observe while the Moon is separated from Praesepe by at least $45^\circ$; and (iii) observe while Praesepe is above the lower elevation limit of Keck I, about $33^\circ$. These observing constraints can be specified with the \texttt{AtNightConstraint}, \texttt{MoonSeparationConstraint}, and \texttt{AltitudeConstraint} objects. Other built-in constraints include: Moon illumination, airmass limits, Sun separation limits (e.g.,~for non-optical observations), and local time constraints. The observing constraint classes take the following parameters as input: targets, times and an observer. The constraints return boolean matrices indicating whether or not those targets are observable at each time. 

\begin{figure}[h]
\centering
\includegraphics[scale=0.7]{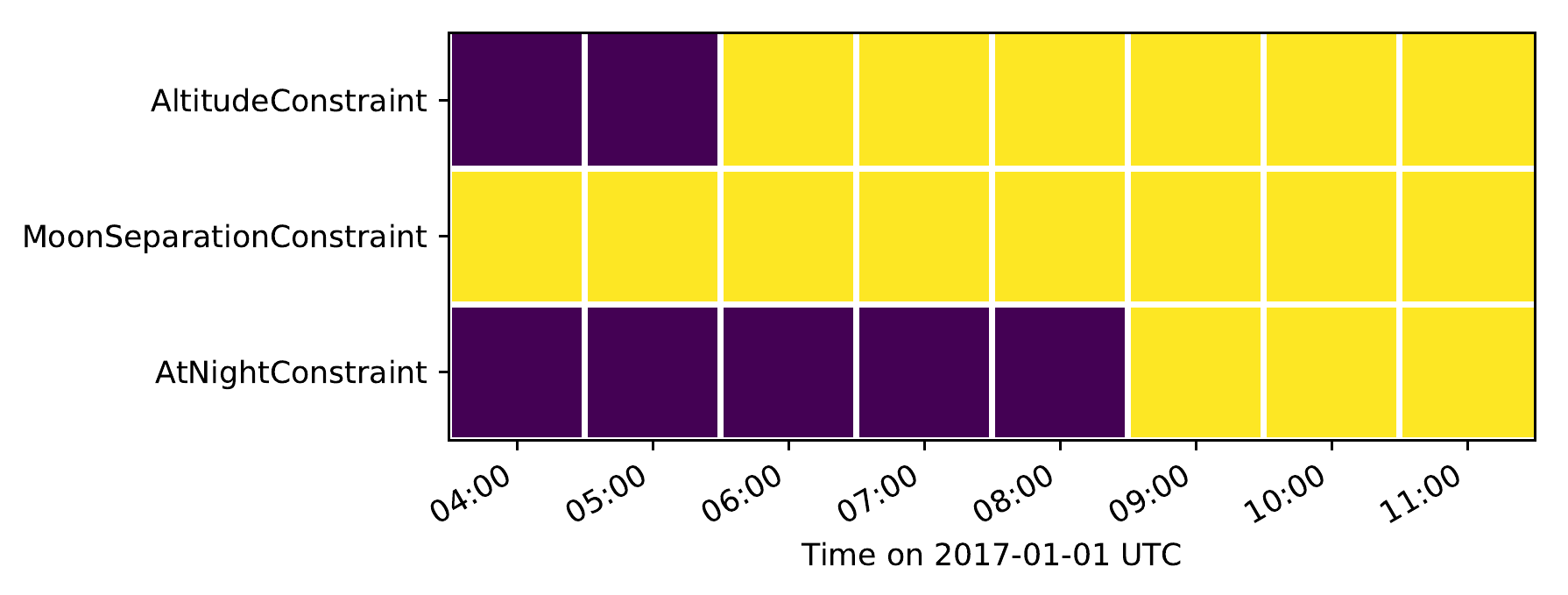}
\caption{Diagram summarizing the ``observability grid'' of Praesepe given each observing constraint, at each hour within the time range. Dark squares represent times when the observing constraint is not satisfied.}
\label{fig:constraints}
\end{figure}

The following code will compute whether or not Praesepe is observable given the constraints listed above. The array \texttt{observablility} will contain \texttt{True} for times when Praesepe is observable given the specified constraints, and \texttt{False} otherwise. We visualize the observability grid in Figure~\ref{fig:constraints}.

A warning may be printed if \astropy or \astroplan need to update the International Earth Rotation and Reference Systems Service (IERS) tables before computing a target's altitude and azimuth. The altitude and azimuth of a target depends on the orientation of the Earth, which varies on short timescales due to shifts in the Earth's moment of inertia. In order to account for these unpredictable variations in the Earth's position with time, astropy (and therefore astroplan) use constantly updated tables from the IERS which specify the Earth's orientation with observations of quasars. 

\begin{lstlisting}
from astroplan import (FixedTarget, Observer, AltitudeConstraint,
                          AtNightConstraint, MoonSeparationConstraint)
from astroplan.utils import time_grid_from_range
from astropy.time import Time
import astropy.units as u
import numpy as np

# Specify observer at Keck Observatory:
keck = Observer.at_site("Keck")

# Use Sesame name resolver to get coordinates for Praesepe:
target = FixedTarget.from_name("Praesepe")

# Define observing constraints:
constraints = [AtNightConstraint.twilight_astronomical(),
                 MoonSeparationConstraint(min=45 * u.deg),
                 AltitudeConstraint(min=33 * u.deg)]

# Define range of times to observe between
start_time = Time("2017-01-01 04:00:01")
end_time = Time("2017-01-01 11:00:01")
time_resolution = 1 * u.hour

# Create grid of times from ``start_time`` to ``end_time``
# with resolution ``time_resolution``
time_grid = time_grid_from_range([start_time, end_time],
                                      time_resolution=time_resolution)

observability_grid = np.zeros((len(constraints), len(time_grid)))

for i, constraint in enumerate(constraints):
    # Evaluate each constraint
    observability_grid[i, :] = constraint(keck, target, times=time_grid)
    
# The plotting commands are omitted from this paper for brevity, and are available in 
# the online documentation. The result is plotted in Figure 2.
\end{lstlisting}

\section{Eclipsing Binary and Transiting Exoplanet Ephemerides} \label{sec:eb}

Suppose you want to observe a newly discovered eclipsing binary, or a well-known transiting exoplanet. You can compute the time of the next primary eclipse or transit event with the \texttt{EclipsingSystem} object. 

\begin{lstlisting}
from astropy.time import Time
import astropy.units as u
from astroplan import EclipsingSystem

# We will compute the next transit/eclipse times relative to this reference time:
reference_time = Time("2017-07-07 00:00")

# Define system parameters:
primary_eclipse_time = Time("2017-07-07 12:00")
orbital_period = 2 * u.day

example_binary = EclipsingSystem(primary_eclipse_time, orbital_period)

print(example_binary.next_primary_eclipse_time(reference_time))
# The printed result is: ['2017-07-07 12:00:00.000']
\end{lstlisting}

With the latest version of astroquery \citep{astroquery}, you can query the NASA Exoplanet Science Institute Exoplanet Archive \citep{Akeson2013} or the Exoplanet Orbit Database \citep{Wright2011, Han2014} for exoplanet system parameters:
\begin{lstlisting}
# You can use astroquery version >= 0.3.7 to get planet parameters from the 
# Exoplanet Orbit Database like this:
from astroquery.exoplanet_orbit_database import ExoplanetOrbitDatabase

planet = ExoplanetOrbitDatabase.query_planet('HD 209458 b')

name = "HD 209458 b"
primary_eclipse_time = Time(planet['T0'], format='jd')
orbital_period = planet['PER']
duration = planet['T14']
eccentricity = planet['ECC']
argument_of_periapsis = planet['OM']

hd_209458_b = EclipsingSystem(primary_eclipse_time, orbital_period,
                                 duration, eccentricity, argument_of_periapsis,
                                 name)

print(hd_209458_b.next_primary_eclipse_time(reference_time))
# The printed result is: ['2017-07-10 01:31:18.521']
\end{lstlisting}

\section{Scheduling Observations} \label{sec:scheduleapp}

In this example, suppose we want to create a schedule for observations at Apache Point Observatory in the first half of the night of 2016 July 7 UTC. We will schedule 16 exposures of Deneb and M13, each in three color filters: B, G and R. We must observe these targets when they meet the following constraints: (1) the airmass of the target is $<3$; (2) the time is between civil twilights; (3) the time is between 02:00-08:00 UTC, which corresponds to the first half of the night at Apache Point.

\astroplan provides control over the many parameters that affect observation scheduling. In the example below, we take into account the slew rate of the telescope, the time it takes to change filters, and a user-input priority for each observing block. 

\begin{lstlisting}
import astropy.units as u
from astropy.time import Time
from astroplan import (Observer, FixedTarget, ObservingBlock, Transitioner,
                          AirmassConstraint, TimeConstraint, AtNightConstraint,
                          PriorityScheduler, Schedule)
from astroplan.plots import plot_schedule_airmass
import matplotlib.pyplot as plt

# Define the observer, at Apache Point Observatory
apo = Observer.at_site("APO")
deneb = FixedTarget.from_name("Deneb")
m13 = FixedTarget.from_name("M13")

# Define start time and end time of the schedule.
noon_before = Time("2016-07-06 19:00")
noon_after = Time("2016-07-07 19:00")

# Specify the read-out time, exposure duration and number of exposures
read_out = 20 * u.second
deneb_exp = 60 * u.second
m13_exp = 100 * u.second
n_exposures = 16   # Number of exposures

# Define the constraints global (constraints on all targets) and specific
# (constraints for individual targets) and make a list of the
# observing blocks that you want to schedule
global_constraints = [AirmassConstraint(max=3, boolean_constraint=False),
                         AtNightConstraint.twilight_civil()]

# This will be the list of observing blocks to schedule:
blocks = []

first_half_night = TimeConstraint(Time("2016-07-07 02:00"),
                                      Time("2016-07-07 08:00"))

for priority, bandpass in enumerate(['B', 'G', 'R']):
    # We want each filter to have its own, independent priority (so that target
    # and reference star are both scheduled)
    b = ObservingBlock.from_exposures(deneb, priority, deneb_exp, n_exposures,
                                          read_out, constraints=[first_half_night],
                                          configuration={'filter': bandpass})
    blocks.append(b)

    b = ObservingBlock.from_exposures(m13, priority, m13_exp, n_exposures,
                                          read_out, constraints=[first_half_night],
                                          configuration={'filter': bandpass})
    blocks.append(b)

# Define how the telescope transitions between the configurations defined in the
# observing blocks (target, filter, instrument, etc.).
filter_change_times = {'filter': {('B', 'G'): 10*u.second,
                                      ('G', 'R'): 10*u.second,
                                      'default': 30*u.second}}

# How fast does the telescope move, on average?
telescope_slew_rate = 0.8 * u.deg/u.second
transitioner = Transitioner(slew_rate=telescope_slew_rate,
                               instrument_reconfig_times=filter_change_times)

# Initialize the scheduler
priority_scheduler = PriorityScheduler(constraints=global_constraints,
                                           observer=apo, transitioner=transitioner)

# Create a Schedule object, which the Scheduler will insert observing blocks
# into
schedule = Schedule(noon_before, noon_after)

# Run the scheduler, and put the observing blocks ``blocks`` into the
# schedule object ``schedule``
priority_scheduler(blocks, schedule)

# To get a plot of the airmass vs where the blocks were scheduled
plt.figure(figsize=(8, 6))
plot_schedule_airmass(schedule)
plt.tight_layout()
plt.legend(loc="upper right")
plt.savefig("schedule.pdf", bbox_inches="tight")
plt.show()

# The resulting plot is shown in Figure 3
\end{lstlisting}

\begin{figure}[h]
\centering
\includegraphics[scale=0.5]{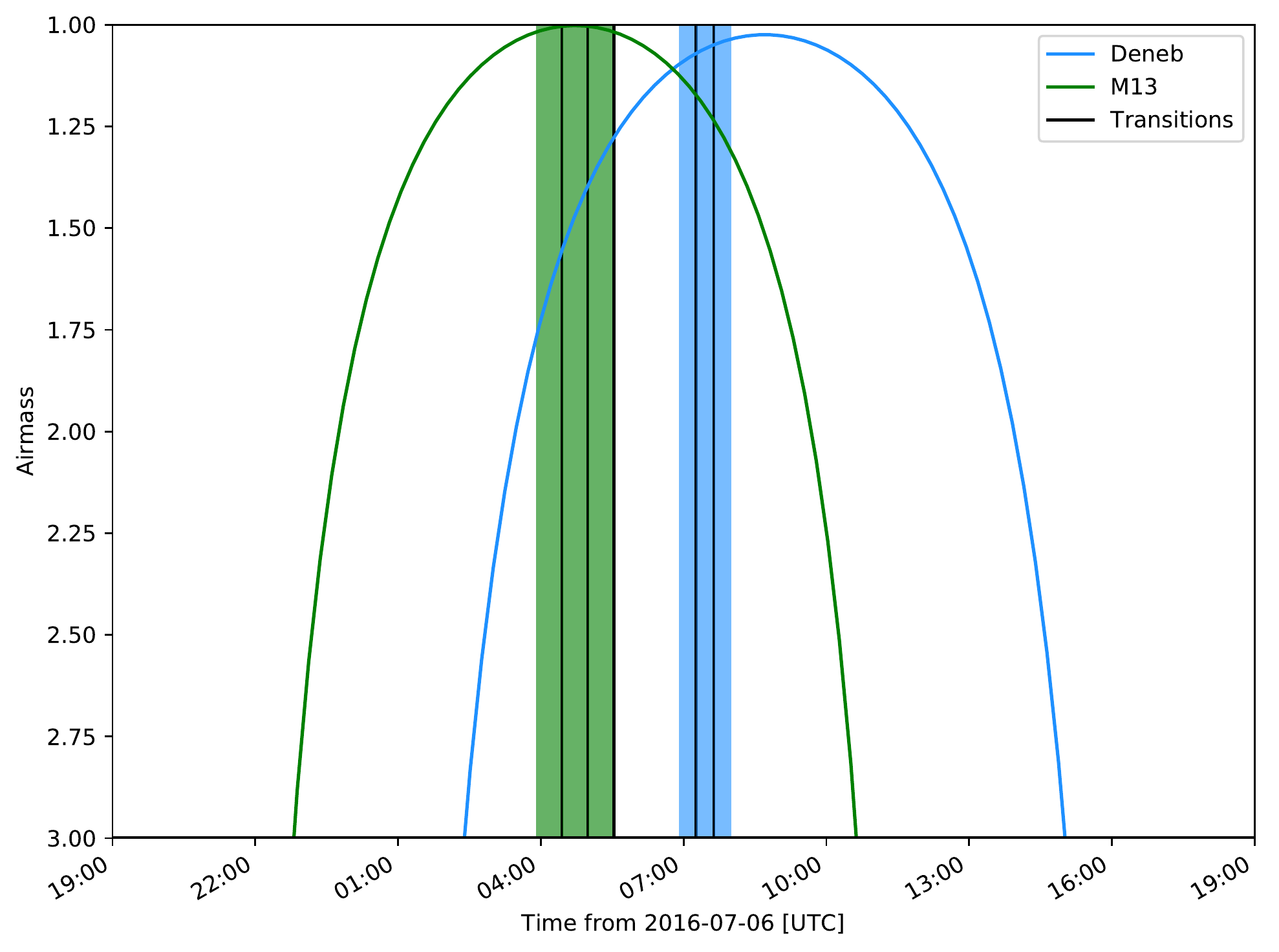}
\caption{Airmass plot showing the scheduled observing blocks. Since we constrained the observations to occur before 8:00 UTC, but Deneb doesn't reach its highest altitude until after 8:00 UTC, the scheduler assigned the Deneb observing blocks as late as possible before the 8:00 UTC, thus minimizing the airmass of Deneb during the observations. Since M13 is observable at its minimum airmass, the scheduler centered the three observing blocks on the times when M13 is at minimum airmass. The black lines between observing blocks represent transitions, which account for instrument reconfiguration dead time --- in this example, filter changes and telescope slews add some dead time.}
\label{fig:schedule}
\end{figure}

\end{document}